\documentclass[screen]{acmart}

\AtBeginDocument{%
  }

\setcopyright{acmlicensed}

\usepackage[utf8]{inputenc}
\usepackage{array}
\usepackage{multirow}
\usepackage{booktabs}
\usepackage{wrapfig}
\usepackage{graphicx} 

\usepackage{ragged2e} 

\usepackage{geometry} 
\usepackage{pgfplots}
\pgfplotsset{compat=1.17} 
\usepackage{caption}
\usepackage{subcaption}
\usepackage{tikz}
\usepackage{arydshln}
\setcopyright{acmlicensed}
\copyrightyear{2025}
\acmYear{2025}
\acmDOI{XXXXXXX.XXXXXXX}
\acmConference[25]{}{November 5–7, 2026}{USA}
\acmISBN{978-1-4503-XXXX-X/2018/06}

\begin{document}

\title[Ghostcraft]{\textit{`Ghostcrafting AI'}: Under the Rug of Platform Labor}

\author{ATM Mizanur Rahman}
\affiliation{
  \institution{University of Illinois Urbana-Champaign}
  \city{Urbana}
  \state{Illinois}
  \country{USA}}
\email{amr12@illinois.edu}

\author{Sharifa Sultana}
\affiliation{
  \institution{University of Illinois Urbana-Champaign}
  \city{Urbana}
  \state{Illinois}
  \country{USA}}
\email{sharifas@illinois.edu}

\renewcommand{\shortauthors}{Rahman et al.}

\begin{abstract}
Platform laborers play an indispensable yet hidden role in building and sustaining AI systems. Drawing on an eight-month ethnography of Bangladesh's platform labor industry and inspired by Gray and Suri, we conceptualize \textit{Ghostcrafting AI} to describe how workers materially enable AI while remaining invisible or erased from recognition. Workers pursue platform labor as a path to prestige and mobility but sustain themselves through resourceful, situated learning—renting cyber-café computers, copying gig templates, following tutorials in unfamiliar languages, and relying on peer networks. At the same time, they face exploitative wages, unreliable payments, biased algorithms, and governance structures that make their labor precarious and invisible. To cope, they develop tactical repertoires such as identity masking, bypassing platform fees, and pirated tools. These practices reveal both AI's dependency on ghostcrafted labor and the urgent need for design, policy, and governance interventions that ensure fairness, recognition, and sustainability in platform futures.


\end{abstract}



\received{20 February 2007}
\received[revised]{12 March 2009}
\received[accepted]{5 June 2009}

\maketitle

\section{Introduction}
The rapid expansion of AI and digital platforms has created new infrastructures for outsourcing tedious yet indispensable forms of labor. Platforms such as Amazon Mechanical Turk and Upwork have become central to the production of data-driven markets, relying on a distributed workforce to perform annotation, moderation, and design tasks that sustain the daily operations of AI systems \cite{irani2015cultural}. These invisible contributions underpin the very possibility of machine learning, yet the workers performing them often face low pay, limited rights, and inadequate recognition \cite{1-silberman2018responsible, 2-irani2015difference, 9-tandon2022barriers}. Prior HCI research has documented how algorithms control workers, how opaque evaluation systems diminish their agency, and how crowd laborers are frequently treated as replaceable within infrastructures of value creation \cite{2-irani2015difference, 12-kaun2020shadows, 22-png2022tensions}. Although scholars have explored avenues for worker organizing, participatory governance, and ethical platform design, systemic inequities persist, leaving the very labor that makes AI possible structurally precarious.

Against this backdrop, we introduce the concept of \textit{``Ghostcrafting AI''} to describe how platform workers in the Global South materially contribute to building and sustaining artificial intelligence systems while remaining hidden, anonymized, or erased from recognition. Ghostcrafting foregrounds both the dependency of AI on these workers' expertise and the structural processes that obscure their authorship. Being motivated by Gray's and Suri's Ghost Work \cite{gray2019ghost}, we focus on Bangladesh because platform labor is not only a growing livelihood strategy but also a national development priority, actively promoted in state policies and supported by global development organizations under agendas of digitalization \cite{worldbank2020edge-press, uncdf2022mou, unbangladesh2022mou}, AI innovation \cite{bangladesh2019aistrategy-draft, undp2025youth-ai}, and the UN Sustainable Development Goals \cite{bdgov2025-sdg-tracker, undp-sdgintegration-bd}. Our eight-month-long ethnographic study employed interviews, biography making, and observations with Bangladeshi platform workers and stakeholders (n=34) to investigate the everyday practices of Bangladeshi platform workers and find how global concerns of unfair pay, biased algorithms, and absent accountability become locally experienced and contested, as well as the contextual challenges the platform workers experience. In this regard, we solicited answers to the following research questions:

\begin{quote}
\textit{RQ1:} How do workers in Bangladesh enter into platform labor, and what personal, social, and infrastructural conditions motivate and shape their trajectories into digital work?\\
\textit{RQ2:} What forms of situated knowledge, skill development, and peer learning emerge as workers navigate global digital marketplaces with inadequate formal training or institutional support? \\
\textit{RQ3:} How do workers experience and negotiate the challenges of unfair pay, algorithmic bias, recognition gaps, and structural discrimination in platform labor? \\
\textit{RQ4:} What design and policy futures can be envisioned to transform platform labor into a more sustainable, equitable, and accountable form of work in the Global South?
\end{quote}

Our findings show that Bangladeshi platform work is often perceived as a prestigious tech job that offers higher social status within low-income and low-education communities. Workers develop expertise through highly situated practices such as renting cyber-café computers, copying gig descriptions from top sellers, following YouTube tutorials in unfamiliar languages, and relying on WhatsApp peer groups—demonstrating how skill-building occurs outside formal training. Yet participants reported how overpriced certification schemes, visibility algorithms that penalize inactivity or rating drops, and exploitative review mechanisms collectively undermine careers, reinforcing inequities and silencing feedback. They also described how their contributions are routinely erased, as clients and platforms impose non-disclosure agreements (NDAs), forbid completed work in portfolios, or threaten legal action for claiming authorship. To survive these structural disadvantages, workers employ tactical repertoires: posing as U.S.-based freelancers via VPNs and borrowed IDs, encoding WhatsApp numbers inside images to bypass platform fees, stabilizing careers with pirated ``safe'' software, and boosting visibility through dozens of low-pay tasks. These insights extend Gray's and Suri's Ghost Work by highlighting Global South–specific structural gaps and survival repertoires, emphasizing communal learning infrastructures as central to sustaining platform labor, and translating these critiques into designable interventions and redistributive obligations for HCI and HAI systems.

Our work makes four key contributions to HCI, HAI, and platform labor scholarship. \textbf{First}, we present an eight-month ethnography of Bangladeshi platform workers, providing biographical accounts, learning trajectories, and survival repertoires that expose the structural inequities and global asymmetries of platform labor in the Global South. \textbf{Second}, we show how precarity, biased algorithms, and erased authorship intersect with communal learning ecologies and tactical survival, reframing platform labor as a socio-technical system of labor rights, governance, and infrastructure rather than isolated tasks. \textbf{Third}, we conceptually extend Gray and Suri's notion of ghost work by introducing ghostcrafting AI, which highlights not only the invisibility of platform labor but also the situated, communal, and tactical practices through which workers materially sustain AI systems. \textbf{Fourth}, we translate these insights into obligations for HCI and HAI design and engage in policy discourse by proposing actionable interventions such as peer learning infrastructures, accountable algorithms, recognition mechanisms, regional training funds, portable credentials, and emergency income supports that can move beyond critique toward structural change.

\section{Related Work}
Previous reviews explored narrower aspects of crowdsourcing—such as its ethical, managerial, or technological dimensions. Recent research has also emphasized the role of stakeholder engagement and participatory practices in designing more inclusive and accountable AI and platform systems, reflecting growing efforts to address ethical and structural gaps \cite{eaamo-kallina2024stakeholder, eaamo-delgado2023participatory, eaamo-sloane2022participation, eaamo-birhane2022power}. Building on this scholarship, our work contributes by studying Bangladeshi platform labor and introducing the concept of ghostcrafting AI.

\subsection{Gig Work, Precarity, and Worker Politics}
Gig work refers to short-term, flexible jobs facilitated through online platforms, where workers are not regular employees and therefore lack access to traditional employment benefits \cite{gig-spreitzer2017alternative, gig-watson2021looking, 66-wu2024gig}. The broader gig economy connects workers with clients for short-term tasks, often promoted as offering flexibility and independence. However, research highlights that this flexibility comes at the cost of job insecurity, unstable income, and an absence of rights protections \cite{gige-donovan2016does, gige-koutsimpogiorgos2020conceptualizing, gige-roy2020future, 58-stewart2017regulating}. Gig workers frequently encounter low wages, scams, and exploitative conditions, with platforms exercising algorithmic management that enforces compliance and reduces autonomy \cite{lit1-hickson2024freedom, lit3-gerber2021community, 10-irani2023algorithms, 15-mcinnis2016taking, 28-ettlinger2016governance, rahman2025digital}. Other studies show how deceptive pricing and identity barriers further reinforce the unequal power between workers and platforms \cite{lit5-grohmann2022platform, rahman2025daiem}. Digital platforms typically privilege efficiency over fairness, which exacerbates existing inequalities and amplifies uncertainty for workers \cite{lit4-kuhn2021human, 5-hansson2016crowd, 27-du2024ethical}.

Despite being framed as empowering, platform labor often conceals domination and precarity \cite{lit1-hickson2024freedom, lit2-hansson2014micro}. Recent scholarship has considered how worker participation and feedback systems might rebalance power, though such mechanisms must avoid becoming extractive or tokenistic \cite{eaamo-sloane2022participation, eaamo-birhane2022power, eaamo-corbett2023power, eaamo-russo2024bridging, 3-salehi2015we, 4-irani2016stories}. Our study builds on these insights by examining how Bangladeshi workers pursue gig work as a prestigious and socially valued alternative to low-income local jobs, yet quickly confront systemic inequities that limit autonomy and reinforce global hierarchies. This literature motivates our first research question on how workers enter platform labor and navigate its socio-political constraints.

\subsection{Learning, Knowledge, and Worker Adaptation}
Prior reviews of crowdsourcing literature have examined ethics, management, and learning opportunities. Durward \cite{136-durward2016there} used the PAPA framework (privacy, accuracy, property, accessibility) to analyze ethical dimensions and identified a lack of attention to workers' experiences and rights. Bhatti's survey of 234 articles proposed a three-step framework—task design, implementation, and aggregation—emphasizing shortcomings in evaluation, incentives, and management \cite{137-bhatti2020general}. Drechsler \cite{138-drechsler2025systematic} systematically reviewed learning in crowdwork, showing that platforms, digital tools, and community networks shape how workers acquire skills. Bazaluk \cite{139-bazaluk2024crowdsourcing} explored the link between migration and crowdsourcing, showing how migrant workers engage with platforms but warning against overlooking long-term implications.

Other research has highlighted how rural and marginalized workers face distinctive challenges in crowdwork. Flores-Saviaga et al. \cite{hcomp-flores2020challenges} showed how rural U.S. workers confronted infrastructural and social barriers, while Abbas and Gadiraju \cite{41-abbas2022goal} examined goal-setting practices that shape worker motivation. Imteyaz et al. \cite{hcomp-imteyaz2024human} further argued for human-centered designs tailored to the needs of diverse populations. Taken together, these studies show that learning is not linear or formalized but mediated by informal infrastructures of sharing, imitation, and peer support \cite{11-mcinnis2016one, 18-feldman2021we}. Our findings extend this literature by showing how Bangladeshi workers learn through resource-constrained and situated practices, including renting cyber-café computers on shifts, copying gig descriptions from successful sellers, following YouTube tutorials in unfamiliar languages, and troubleshooting problems in WhatsApp peer groups. These practices highlight how global knowledge is adapted under local constraints, complicating mainstream narratives of digital ``upskilling." This literature informs our second research question on how Bangladeshi workers cultivate situated expertise and adapt through informal infrastructures.

\subsection{Fairness, Algorithmic Bias, and Invisible Labor}
Another major strand of research highlights issues of fairness, pay, and algorithmic bias in digital labor. Workers in economically vulnerable regions are often paid below minimum wage and remain unrecognized for their contributions \cite{10-irani2023algorithms, 109-haralabopoulos2019paid, 55-salminen2023fair}. Task-based payment systems exacerbate precarity by ignoring unpaid labor, such as waiting for tasks or investing time to learn new tools \cite{111-berg2015income, 112-chen2024we}. Studies also show how algorithms assign tasks in biased ways, privileging certain regions or profiles \cite{117-zhang2017consensus, 119-eickhoff2018cognitive}. Further, limited worker control over task structure reduces their agency and amplifies exclusion \cite{13-grohmann2021beyond, 116-ford2015crowdsourcing, 15-mcinnis2016taking}.

Beyond pay and workflow, research has increasingly highlighted the invisibility of labor contributions. Li et al. \cite{faact-li2023dimensions} and Whitney \cite{faact-whitney2024real} argue that data labor and synthetic data practices appropriate contributions without clear consent or benefit sharing, deepening regulatory and ethical gaps. Our findings confirm and extend this by showing how Bangladeshi workers' authorship is systematically erased: they are compelled to sign non-disclosure agreements, prevented from displaying completed projects in portfolios, and sometimes threatened with legal consequences for claiming credit. These experiences echo Gray and Suri's account of ghost work but extend it by foregrounding how invisibility is coupled with biased algorithms and exploitative ratings that perpetuate long-term precarity \cite{12-kaun2020shadows, 14-willcocks2019hidden}. This literature grounds our third research question on how Bangladeshi workers experience unfairness, bias, and structural invisibility in platform economies.

\subsection{Accountability, Sustainability, and the Future of Work}
Research also highlights the absence of accountability in platform governance. Workers report sudden suspensions, underpayment, and unanswered grievances \cite{120-grossman2018crowdsourcing, 121-zou2018proof, 16-mcinnis2016running}. Platforms rarely face consequences because no independent oversight structures exist \cite{13-grohmann2021beyond}. While guidelines for audits, transparency, and feedback systems have been proposed, they are rarely adopted in practice \cite{16-mcinnis2016running, 17-mcinnis2017crowdsourcing, eaamo-barker2023feedbacklogs}. This absence of accountability reinforces power imbalances, leaving workers fearful of speaking out and skeptical that reforms will address their concerns.

Sustainability is another critical gap. Most crowdwork is organized around short-term, repetitive tasks with little opportunity for training or career development \cite{17-mcinnis2017crowdsourcing, 25-naude2022crowdsourcing, 109-haralabopoulos2019paid}. As a result, workers experience unstable income, fatigue, and burnout, rendering digital labor unsustainable in the long term \cite{13-grohmann2021beyond, 46-de2022understanding, 111-berg2015income, 55-salminen2023fair}. Widder \cite{faact-widder2024epistemic} argues that fairness, sustainability, and accountability are rarely studied together, obscuring their interdependencies. Our findings contribute by showing how Bangladeshi workers create tactical repertoires—posing as Western freelancers, embedding WhatsApp numbers in images to avoid fees, stabilizing careers with pirated ``safe" software, and boosting ratings through low-paying tasks—as improvised infrastructures of survival. These repertoires highlight both the absence of institutional support and the resilience of workers who must build sustainability through fragile, self-made tactics. This literature informs our fourth research question and contribution: tracing survival strategies and envisioning design and policy futures for more sustainable platform economies.

\section{Methods}
Our eight-month-long ethnography took place during May 2024 - Aug 2024; Dec 2024 - Jan 2025, and May 2025 - Aug 2025 in seven sites in Dhaka (one in each Mirpur, Farmgate, Rampura, Sher-e-Bangla Nagar, and two in South Banasree) and Sadar in Jashore. Seven organizations helped us conduct the research: WAN Computer Training Center, Ayash Trade Computer Training Center, Ambition Shorthand Training Center, Job house Shorthand and Computer Training Center, Ahad Computer Training Institute, Job Care Computer Training Institute in Dhaka, and Rural Reconstruction Foundation in Jashore. We employed observation, interviews, and biography making to gather in-depth insights into Bangladeshi platform labor for AI.

\begin{wraptable}{R}{0.51\textwidth}
\vspace{-30pt}
\begin{center}
\begin{tabular}{|rl|}
\hline
Total Participants: & 34 (Female: 6, Male: 28)\\
\hdashline
\multicolumn{2}{|c|}{\textbf{Participant Roles (multiple possible)}} \\
\hdashline
Data Annotation: & 15 (Female: 4, Male: 11)\\
Graphic Design / UI-UX: & 14 (Female: 1, Male: 13)\\
Content Writing / Translation: & 10 (Female: 4, Male: 6)\\
Web Development: & 14 (Female: 0, Male: 14)\\
Data Entry: & 8 (Female: 3, Male: 5)\\
Trainers: & 7 (Female: 1, Male: 6)\\
Other Roles (e.g., support, marketing): & 9 (Female: 2, Male: 7)\\
\hdashline

\multicolumn{2}{|c|}{\textbf{Platforms (multiple possible)}} \\ 
\hdashline
Upwork: & 24 (Female: 3, Male: 21)\\
Fiverr: & 23 (Female: 5, Male: 18)\\
Freelancer.com: & 15 (Female: 4, Male: 11)\\
Crowdgen: & 10 (Female: 1, Male: 9)\\
99designs: & 9 (Female: 1, Male: 8)\\
Toloka: & 6 (Female: 2, Male: 4)\\
PeoplePerHour: & 5 (Female: 0, Male: 5)\\
\hdashline

\multicolumn{2}{|c|}{\textbf{Age Range (in Years)}} \\
\hdashline
All: & 19--39, median 27\\
Male: & 19--39, median 27\\
Female: & 21--34, median 26\\
\hdashline

\multicolumn{2}{|c|}{\textbf{Education}} \\ 
\hdashline
Primary Education: & 9 (Female: 3, Male: 6)\\
Secondary School Certificate: & 8 (Female: 1, Male: 7)\\
Higher Secondary School: & 9 (Female: 1, Male: 8)\\
Bachelor’s Degree: & 6 (Female: 1, Male: 5)\\
Master’s Degree: & 2 (Female: 0, Male: 2)\\
\hline

\end{tabular}
\end{center}
\caption{The demographic details of the participants}
\vspace{-35pt}
\label{Tab:demo}
\end{wraptable}

\subsection{Participant Recruitment}
We recruited participants by distributing flyers and through NGOs. We reached out to the NGO and training institutes, explaining the research, and sought opportunities to observe their training classes and engage with the trainers and trainees. Some of them expressed interest in participating in interviews and biography making. We also recruited interview participants from the platform labor community by posting the flyer for our research on social media. Also, often participants referred others who might be a good fit for the study \cite{goodman1961snowball}. All the participants were 18 years or older with at least six months of prior experience in platform work (e.g., data cleaning, content moderation, annotation, image labeling, etc.). See Table~\ref{Tab:demo} for participants' demographics. 

\subsection{Observation}
\begin{figure}[t] 
  \centering
  \includegraphics[width=0.8\linewidth]{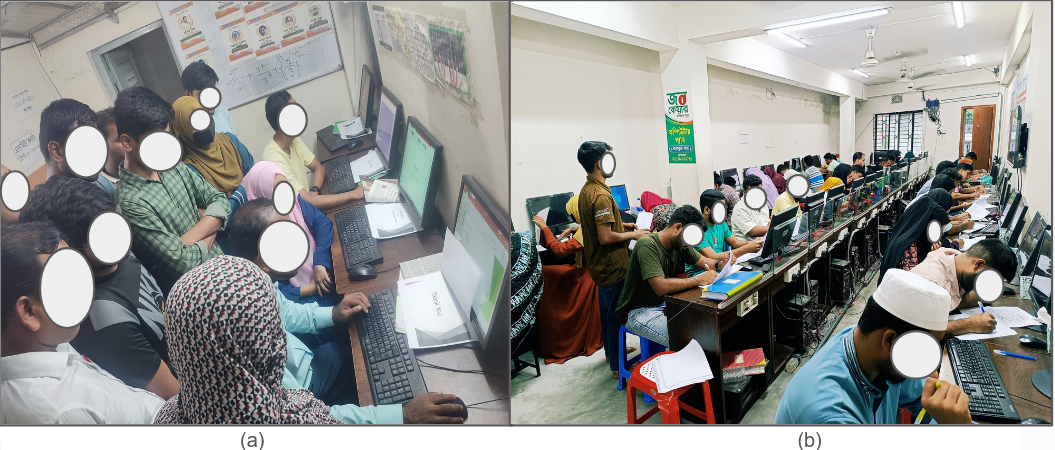} 
  \vspace{-5pt}
  \caption{(a) A training session at one of the training centers. The trainer (in a light blue shirt) sits at one computer and teaches, while the students gather around closely to learn from him. (b) A handwritten exam is taking place at the training center. The person standing with papers is an institute staff member making sure that no one copies from others.}
  \label{fig:platform}
\end{figure}

All these training centers and RRF have a long history of running training programs under government and international NGO initiatives. Hence, they facilitated our observation sessions upon our request. The coordinators of the programs scheduled the sessions for us and took us to the training classes to introduce us to them. Upon our explaining the goal of this work, the participants asked for clarification before they agreed to be observed. We conducted 18 three-to-four-hour-long observation sessions (total almost 60 hrs) to understand their training processes, the materials they emphasize, and their shared insights from experiences. We took detailed notes in our notebook during the sessions and took photos of the classroom with the participants' permission. (See Figure ~\ref{fig:platform}) 

\subsection{Semi-structure Interviews}
We conducted 34 semi-structured interviews \cite{ruslin2022semi}. All interviews were conducted in Bengali, the native language of both the participants and the researchers. At the beginning of each interview, we explained the purpose of the study and sought informed consent. During the interviews, we During the interviews, we asked participants about their entry into platform work and the social conditions that shaped these trajectories, the ways they built and shared skills, their experiences of unfair pay, bias, and invisibility, and the tactics they developed to navigate or resist these challenges. Each session lasted between 45 to 60 minutes and was scheduled according to the participant’s availability. Interviews were held either remotely over Zoom (12) or in person, depending on what each participant preferred. We took detailed notes. 32 participants gave us permission to audio-record their sessions. These materials were later transcribed and translated for open coding and thematic analysis \cite{fereday2006demonstrating, boyatzis1998transforming}. 

\subsection{Biography Making}
We developed three detailed biographical sketches from the participants. These accounts draw on extended interactions with a subset of participants and capture their family situations, educational paths, and struggles alongside their work. Rather than complete life histories, they are ethnographic portraits that illustrate diverse trajectories into platform labor and the meanings participants attach to their work. We selected three individuals to reflect different entry points and experiences within Bangladesh's platform economy: one became a trainer after years of freelancing, another turned to platform work after leaving formal education, and a third combined multiple forms of digital work for survival. Following traditions of narrative analysis \cite{linde1993life,weiss1995learning}, these biographical accounts highlight both the situated skill-building of workers and the broader social processes through which livelihoods and identities are shaped in platform labor.

\subsection{Data Collection and Analysis}
We collected 160 pages of fieldnotes, more than 200 images, and 22 hours of audio recordings. We transcribed the audio recordings and translated them from Bengali to English, removed identifiers before conducting an open coding and thematic analysis \cite{fereday2006demonstrating, boyatzis1998transforming}. Two members of the research team independently reviewed the transcripts to become familiar with the content. During the open coding process, we allowed codes to emerge inductively, capturing participants’ reflections, stories, and critiques of their jobs and the ecosystem. The initial codes included topics such as fairness in task distribution, lack of clarity in evaluation systems, challenges with identity verification, and issues with payment mechanisms. We clustered them into themes through multiple iterations (see codebook included in the supplementary materials). These themes are presented in the next section.

\subsection{IRB and Positionality} 
This study was reviewed and approved by our university's Institutional Review Board (IRB). All participants provided informed consent and were assured confidentiality, with pseudonyms used throughout. As researchers, our positionality shaped the study: all the authors are Bangladeshi and shared linguistic and cultural familiarity with participants, which helped establish trust. We acknowledge that our academic and socioeconomic positions differ from those of participants, and we sought to mitigate these imbalances through long-term engagement and reflexive analysis.
\section{Biographies}
We start our findings by presenting three biographical sketches of platform workers. These life histories illustrate how workers enter platform labor, build expertise through situated learning, navigate its social-emotional strains, and negotiate futures under systemic uncertainty. The names used are pseudonyms. 

\subsection{Persona-1: Habib}
Habib is a 38-year-old platform worker and trainer in Khilgaon, Dhaka. He was born and raised in a lower-middle-class family and attended school and higher secondary school in Kushtia, a suburban town in western Bangladesh. He traveled to Dhaka for university admission exams, but failed. He returned home disappointed, yet his family urged him to finish his education and find a stable job. Habib enrolled in a degree college near home, but his path soon took a different turn. Around 2011, while still in degree college, he stumbled upon a section in \textit{Computer Jagat} magazine that changed his life. A column by Zakaria Chowdhury introduced him to this job, where people could earn online using just a computer and an internet connection. The articles offered step-by-step instructions, but he could not start them because he lacked both.  

Curiosity and financial needs drove Habib to find training centers in Dhaka through ads. His family opposed leaving education, but he moved to Dhaka anyway, following two childhood friends with similar interests. They shared a small rented room in Farmgate, Dhaka, and enrolled in a six-month course program. Their hopes were quickly dashed, as the center turned out to be a scam. It vanished when Habib went home for Eid vacation. However, he still did not give up. Through a classmate from the scam training center, Habib found a mentor who was already earning online. He and his friend visited the mentor's home daily, paying to learn directly. In several months, he learned the basics of online work like data entry, Microsoft Excel, bidding on \textit{oDesk} (now \textit{Upwork}), and client communication. Without a personal computer, training was difficult.  

Habib practiced and worked at local cyber cafes. Every day, he booked a computer for 10 to 12 hours. As a regular customer, he received discounts from the owner. He built his this job career from scratch there. Internet data costs were high, and connections came through a USB modem, so he avoided video or audio calls with clients. He worked with intense focus, making the most of limited resources. Habib's efforts paid off. He began receiving regular tasks and building trusted relationships with clients. As his reputation grew, others at the cafe asked for help. This sparked an idea: Habib could teach while he worked. He started offering informal lessons to newcomers. Over the years, this turned into a full-time profession. Now he runs a small training center in Khilgaon, teaching data entry, UI/UX design, web development, resume building, and communication. He also guides students on portfolios and using LinkedIn effectively.  

Habib is gradually shifting toward team-based work now. Five people currently work under him, and he plans to expand by delegating tasks while focusing on mentoring and training. The income from both allows him to support his family. He sees a bright future but does not sugarcoat the challenges. According to him,  

\begin{quote}
\textit{``This field is good, but it is not easy. You will struggle for the first 5-6 years. If you survive those early years, you will do fine. But most people drop out before that.''}
\end{quote}

He explains that this job is both mentally and physically demanding. It requires long hours, quick typing, fast computer navigation, and constant adaptation. He emphasizes the importance of building strong client relationships and peer support systems through online communities like Facebook groups, forums, and knowledge-sharing spaces. Although peer spaces were once essential, he notes that now, with more resources online, newer folks rely less on peers and more on \textit{YouTube, Reddit, Google}, and even \textit{ChatGPT}. For him, the biggest challenge today is market saturation, which is making it harder to secure consistent work even for skilled professionals. Another issue is monotony. Habib finds this job repetitive and isolating at times, especially compared to office jobs with face-to-face interaction. Yet the freedom and potential outweigh the drawbacks. Despite everything, Habib remains optimistic. He believes that anyone with dedication, patience, and hunger to learn can build a sustainable career online. He hopes his children consider this path, too, though only if they are truly interested.

\subsection{Persona-2: Shamim}
Shamim is a 27-year-old platform worker who works from his home in Dhaka. Born into a middle-class family, he grew up watching movies and gaming on computers. From an early age, he heard about earning money with computers but never explored it seriously. Like many Bangladeshi households, his family saw Computer Science as a prestigious career. Shamim followed their wishes and enrolled in a university to study Computer Science and Engineering. Soon, however, he realized coding and programming did not suit him. The coursework felt difficult and uninspiring. A year into his degree, his father passed away, worsening the family's financial situation. Rising education costs and his lack of interest led him to drop out of university and search for alternatives.

Uncertain about his next steps, Shamim considered tutoring or a job until a senior introduced him to online work. Around the same time, he saw ads for training centers offering short courses in graphic design, UI/UX, data entry, and related skills. Unlike programming, these felt accessible and creatively engaging. He enrolled in a graphic design program, where he learned design tools, visual principles, and basic web development. More importantly, he gained confidence in communication. While technical skills came quickly, he recalls that the training center's most lasting impact was improving his soft skills.

\begin{quote}
\textit{``My trainers told me that if you want to succeed in this field, communication is the key. You need to build a good relationship with the client you are working for, because if the client trusts you, they will come back with more work. So your English-speaking skills, communication, and ability to maintain client relationships are really important in this domain.''}
\end{quote}

Following this advice, Shamim took a short English course to strengthen client communication. Over time, he became skilled at writing professional messages, negotiating, and presenting himself effectively. He credits his success to two core strengths: communication and project management. Today, he works full-time from home with his own computer and internet connection. Yet his relationship with this work remains complicated, and he often questions its sustainability.

\begin{quote}
\textit{``I have been doing this work for a few years, but my situation has not improved much. I try to contribute to my family, but I do not think the amount is enough. I have not married yet because of these financial issues, and I hope to settle down only when I become financially stable. Sometimes I do feel doubtful about whether I chose the right profession or if I should have followed a more traditional career path with a regular 9-to-5 job instead.''}
\end{quote}

Shamim is skeptical about long-term prospects. While low-paying tasks like image labeling and data entry are common, they provide little income. Higher-quality jobs require strong client ties or specialized skills, making this work unreliable as a full-time career. He would not strongly recommend it to others, except as part-time work. According to him, it might suit housewives or students, but for full-time workers, jobs are inconsistent. Despite his doubts, Shamim continues to learn and adapt. He stays connected with peers through a \textit{WhatsApp} group formed during training, where members share resources, troubleshoot issues, and discuss emerging skills.

\begin{quote}
\textit{``During the training, I met some people with whom I formed a WhatsApp group to communicate and support each other. If anyone gets stuck on a problem, like facing an error they cannot solve, others try to give suggestions, and most of the time, it works. Since this field is evolving rapidly, we also discuss new skills and what we should learn next for our improvement. These small things help us grow together.''}
\end{quote}

Shamim emphasizes that skill-building is a continuous and collective process. Mistakes are natural, and growth comes through doing, failing, and sharing. Yet newcomers often struggle to find direction, overwhelmed by countless videos and courses without clear guidance on which skills matter. Alongside technical hurdles, he faces social barriers. Many people around him, including some family members, do not see his work as real or meaningful. Unlike office jobs, platform work lacks visible structure and recognition, which he finds frustrating. Still, he values the flexibility, autonomy, and freedom from local job markets, where nepotism often limits opportunities. Looking ahead, he hopes to earn more, strengthen his reputation, and eventually start a small company where he can lead a team and grow beyond individual work.

\subsection{Persona-3: Abdullah}
Abdullah is a 29-year-old crowdsourcing worker living in Dhaka. He was born into a lower-middle-class family in Satkhira, where his father farmed to support the household. Despite financial hardship, Abdullah excelled in school and first discovered computers in his Higher Secondary School's small lab with three machines and unstable internet. Weekly ICT classes introduced him to programming and HTML, and he vividly recalls the excitement of finally practicing on a computer, creating database tables and editing rows. Inspired, he told his teacher he wanted to study Computer Science in Dhaka. Although encouraged, his family could not afford further education after his HSC, and he was pressured to earn instead. He took a job at a clothing store in Satkhira, earning only BDT. 12000 (USD. 100) monthly, with little hope of growth.  

\begin{quote}
\textit{``Even if I stayed there for the next five years, my salary would not go above BDT. 25000 (USD. 205). I saw no real future for me, so I started thinking about a better career path. From Facebook advertisements and friends, I heard that people were earning money in online work. I needed computer training. That is when I decided to leave the job and head to Dhaka to join a computer training center.''}
\end{quote}

He moved to Dhaka, enrolled in a training center, and began with Microsoft Office before completing a UI/UX design course. Over a year, he mastered Figma and basic web design, borrowing from family to cover costs. Instructors emphasized efficient problem-solving and networking, encouraging him to use LinkedIn for guidance. After training, Abdullah started doing data entry, annotation, logo design, and micro-jobs. Without a personal computer, he has rented a desk at the training center for three years, working up to eight hours a day.  

Despite his persistence, Abdullah's income remains low. To supplement it, he works part-time at a fuel station, earning BDT. 11000 (USD. 90) monthly. Married with a young child in Satkhira, he dreams of bringing his family to Dhaka but struggles financially. He finds resources scattered and training quality inconsistent, while most available jobs are low-paying. Weak English remains a barrier, and he sometimes relies on ChatGPT to draft client messages. Infrastructure problems further constrain him: unreliable electricity and internet at the center prevent efficiency, and conditions are worse in Satkhira. Still, Abdullah remains determined. He hopes to one day own a computer, work from home, and form a small team. Reflecting on his journey, he notes that he once knew nothing about computers but now can design, build, and earn, even if modestly. He longs to lead his own team.

\subsection{Takeaways from the Biographies}
The biographies portray what shapes the path into platform labor in Bangladesh. \textbf{First}, we show how workers enter the market through disrupted education, economic pressures, or aspirations for prestige, as in Habib's persistence after failed university admission and Abdullah's move from retail work to training centers. \textbf{Second}, we found that expertise emerges through situated learning ecologies, where skills are built not in formal pipelines but in patchwork infrastructures such as 12-hour shifts at cyber-cafés, YouTube tutorials in unfamiliar languages, and WhatsApp peer groups that scaffold everyday troubleshooting. \textbf{Third}, we show how platform labor carries profound social-emotional dynamics: Shamim's doubts about sustainability, Abdullah's struggles with family recognition, and Habib's oscillation between optimism and isolation reveal how digital belonging is fragile, stigmatized, and constantly negotiated. \textbf{Finally}, we showed the precarity in the market's future: while tactical repertoires sustain careers in the short term, aspirations for stability, growth, or team leadership are consistently undermined by systemic precarity.

\subsection{Knowledge, Skills, and Learning Processes}
Success in this work depends on how workers learn, adapt, and connect. Amid shifting client needs and competitive marketplaces, they must update skills, rely on peers, and manage platform relationships. We group these learning processes into three domains: technical skills for task execution, adaptive skills for growth, and relational skills for sustaining client trust and visibility.

\subsubsection{Technical and Tool-Based Skills}
Technical and tool-based skills form the foundation of platform labor. These include functional abilities such as UI/UX design, data entry, logo creation, website layout, and annotation. Workers often start with short courses at local IT centers, government programs, or low-cost institutes, covering tools like \textit{Excel, WordPress, Figma, HTML/CSS,} and annotation software. Abdullah began with basic computer literacy before moving to UI/UX design, while Habib learned \textit{Excel} and \textit{Odesk} navigation from a mentor before advancing independently. Shamim stressed that training centers provided the fastest route to technical and professional growth, especially compared to the steep curve of university programming courses.

Alongside formal courses, most workers engaged in extensive \textbf{informal learning}. \textit{YouTube} tutorials, \textit{Facebook} groups, and pirated courses were the most common sources. Six participants even followed tutorials in Chinese or Russian by imitating visual cues without understanding the language. Eight others mimicked gig descriptions or project samples from successful workers, aligning their profiles with market expectations. As one worker explained:  

\begin{quote}
\textit{``I look for profiles of top sellers and copy their gig descriptions with the same structure and keywords, changing details to match my own work. I also design sample images close to theirs so clients see I can do the same type of job.'' (P10)}
\end{quote}

Workers also improved by observing client examples and adapting styles, such as replicating Excel formats or annotation layouts. Many emphasized micro-skills like fast typing, accurate formatting, and attention to detail, which determined whether deadlines and accuracy standards could be met. Beyond tasks, workers needed to master platform navigation—understanding bidding, ranking, avoiding scams, and presenting work effectively. Here, \textbf{peer consultancy} was crucial. One participant recalled:  

\begin{quote}
\textit{``When I got my first foreign client, I did not know how to reply. I sent the draft to a senior friend, and he rewrote it in a professional way. The client accepted my offer the same day.'' (P19)}
\end{quote}

However, advice could backfire, such as being told to bid at unsustainably low rates. Recently, some workers also experimented with AI tools like \textit{ChatGPT} for troubleshooting or platform updates. These were often useful but not always reliable, especially when instructions did not match software versions. When formal credentials are inaccessible or overpriced, workers lean on informal proofs of competence—samples, templates, and peer endorsements—further elevating the role of informal learning ecologies. Collectively, technical skills were pieced together from courses, tutorials, peers, and trial-and-error, constantly reassembled to keep up with client and platform demands.  

\subsubsection{Experiential and Adaptive Skills}
Participants emphasized the importance of adaptability—learning to self-correct, problem-solve, and grow in uncertain conditions. These skills were rarely taught formally but developed through everyday setbacks such as confusing instructions, failed uploads, or sudden platform changes. For Bangladeshi workers, adaptability became a survival skill refined through repetition and persistence. Early experiences were often marked by confusion and mistakes. Habib, for example, initially struggled with formatting errors and slow submissions at a cyber cafe, but gradually developed strategies like pre-reading guidelines and checking outputs carefully. Many participants said they only improved precision and speed through cycles of repetition and self-monitoring. Some used timers, productivity apps, or personal routines to increase efficiency. Over time, repetition turned unfamiliar tools into intuitive ones, while error-tracking built confidence that persistence would lead to mastery.

Adaptability also involved \textbf{staying responsive} to shifting client demands and platform dynamics. Shamim, who dropped out of a computer science program, retrained himself when demand shifted from logo design to UI/UX and frontend work, relying on online tutorials and template replication rather than formal courses. Abdullah, working with limited access to computers and slow internet, adapted by working late at night when fewer people strained the training center's network. These adjustments required both technical flexibility and mental resilience. Workers also highlighted the role of \textbf{self-management}. Many juggled platform work with part-time jobs or family duties, requiring discipline to maintain focus and manage unpredictable workloads. Habib, having experienced burnout, learned to break large tasks into smaller segments and decline overly demanding projects. Others spoke of adopting informal schedules or productivity tips shared by peers.  Adaptability was thus not simply about acquiring new tools but about building judgment, resilience, and professional maturity. Participants described their evolution from hesitant task-doers to confident problem-solvers who could anticipate issues, improvise fixes, and sustain themselves in precarious conditions.

\subsubsection{Relational and Communication Skills}
Finally, workers consistently emphasized the importance of relational and communication skills. These included learning how to interact with clients, maintain trust, and navigate the informal norms of platforms. Communication needed to be timely, respectful, and culturally appropriate. As Shamim put it, technical skills alone were not enough without the ability to secure and retain client relationships.  

\begin{quote}
\textit{``My trainers told me that if you want to succeed in this field, communication is the key. If the client trusts you, they will come back with more work. English, communication, and the ability to maintain relationships are really important.'' (Shamim)}
\end{quote}

For newcomers like Shamim and Abdullah, communication was the hardest part. Some took short English courses, while others studied sample proposals or used AI tools like \textit{ChatGPT} to reframe unclear messages. Workers often developed templates for common client interactions, browsed forums for phrasing ideas, or saved examples of successful proposals. These practices demonstrate how communication was treated as a learned, strategic survival tool.

\textbf{Peer and community spaces} also played a major role. Workers relied on \textit{Facebook groups, WhatsApp chats, and YouTube channels} to exchange sample messages, troubleshoot issues, or observe how others negotiated with clients. Participants consistently said such peer learning was more practical than official documentation, which felt disconnected from their realities. \textbf{Reputation management} was another critical dimension of relational work. Maintaining strong reviews, avoiding violations, and demonstrating reliability required ongoing effort. A single late submission or poor review could damage visibility, while recovery often took weeks of additional work. Some workers even managed small teams under their accounts, bearing responsibility for both output quality and professional reputation. \textbf{Offline ties} also mattered. Habib maintained client referrals through local companies, Shamim relied on alumni networks for support, and Abdullah aspired to form his own team. These cases illustrate how online and offline relationships intertwined to shape careers. Communication and relationship-building were not natural traits but carefully cultivated practices essential to survival in platform economies. These relational practices are also constrained by attribution limits; NDAs and portfolio bans often prevent workers from showcasing completed work, slowing reputation-building and reinforcing invisibility. Workers recognized that being seen, trusted, and invited back mattered as much as technical skill.  

\section{Challenges and Workarounds in Bangladeshi Platform Labor}
Apart from manageable skill and knowledge level challenges, our participants also noted a number of broader challenges associated with the platforms, policies, and the ecosystem. Below, we describe them:

\subsection{Frequent Traditional Challenges}

\subsubsection{Unfair Competition and Nonuniform Opportunities}
One of the major concerns regarding competition and opportunities \textbf{within platform-level} that the participants noted was about the advantage enjoyed by platform-sponsored course certificate holders; 13 participants perceived this as discriminatory against skilled workers trained elsewhere. P3, P9, and P10 felt the tests were disconnected from real work, suggesting external training could be more effective. Participants also noted that platform courses are overpriced and only benefit the platforms themselves. However, five participants fell prey to scam training programs and opined that it is safer to start with the overpriced, inadequate certification from the platforms. Among the \textbf{beyond platform-level struggles}, varying workload and income from month to month, causing challenges to financial stability, was most frequently noted by at least 14 participants. Additionally, the rise of AI-based automation has led many clients to turn to automated tools instead of hiring human workers, which has reduced the volume of available work, as P5 mentioned. 



\subsubsection{On-the-go Nature: The Last-Message-Rule}
Seventeen participants described how platform structures create constant urgency and pressure, leaving little room for flexibility or self-paced work. From strict deadlines set by clients to rules about platform activity and response times, workers feel pressured to be always on call. This affects their mental health and work-life balance. For example, P11 explained that the clients dictate the deadline and often push them to meet deadlines that are half the duration of the original, anticipated timeline. P3 shared his experience of having his ranking dropped because of his absence due to vacation with family. P18 shared how specific platform rules about communication create anxiety and force hyper-responsiveness. Therefore, if a client sends a message, the seller must respond within an hour, and the final message must always come from the worker --- this is named the \textbf{Last-Message-Rule}. He explained that this is well-established in the laborer community, 

\begin{quote}
\textit{``... (W)e are taught that the last message in a chat must always be from the worker. If a client sends a message, I feel pressure to reply within an hour. And no matter what, the conversation is expected to end with my response. It is hard to relax.", (P18)}
\end{quote}

P20 discussed how this is particularly problematic for cross-border contracts. He was warned for taking a short and announced break, as clients in the U.S. follow different holidays than his, and did not acknowledge his day off for a religious festival. P5 also described the struggle of mismatching time zones with international clients. Most of her clients are USA- or UK-based, forcing her to stay awake at odd hours to maintain responsiveness. She shared that clients sometimes leave negative reviews if she does not respond immediately.


\subsubsection{Language Barriers}
Six participants shared how limited English proficiency or the absence of translation tools often led to lost opportunities and reduced client communication. P2 reminisced about losing clients for poor English and difficult communication early in his career. He eventually took private English tuition. However, P15 and P13 described that the problem multiplies when both parties do not speak English or any other common language. Drawing on his experience with a German client, P13 suggested,

\begin{quote}
\textit{``If there were any translation features, such as automatic chat translation from German to English, or if a human mediator were available to help with translations during meetings, I could have taken on those projects. '' (P13})
\end{quote}

Some participants tried using external automatic translation tools, but they are unreliable. Thus, for Global South workers, language barriers and time zone differences on digital labor platforms create structural disadvantages and lead to miscommunication and exclusion that undermine the platforms' perceived borderless nature.


\subsubsection{Rating and Review Abuse}
Participants were skeptical of platform algorithms, which operate behind the scenes. Twelve participants expressed that client reviews often serve as tools of control. Platforms give excessive weight to client feedback, even when it is subjective or inconsistent. Workers feel their professional standing is fragile and that a single negative review can reduce visibility, job offers, or even lead to restrictions. P4 explained,

\begin{quote}
\textit{``Clients think that giving 4-star reviews is a good review, but on Fiverr, anything less than five stars actually hurts us a lot. Sometimes they even write great feedback in words, but leave just two or three stars. That mismatch creates serious problems for us, because the system only counts the stars, not the written feedback.", (P4)}
\end{quote}

However, P6 noted that poor reviews are sometimes intentional and result from workers refusing to meet extra demands. P10 added that although platforms technically allow sellers to review clients, this has little impact. Thirteen participants emphasized that only clients' opinions matter during and after projects. Even when workers follow instructions and deliver high-quality outputs, clients frequently request repeated changes or use reviews to maintain control. P12 described how clients extend project scope without compensation. Workers cannot say no for fear of poor reviews. P16 explained that even after work is completed, clients may still give negative reviews arbitrarily, damaging visibility and reducing opportunities. Hence, Nine participants emphasized that platforms prioritize client satisfaction over worker protection. As P23 explained,

 \begin{quote}
\textit{``At the end of the day, clients bring money to the platform and workers. So platforms will always listen to them first because platforms are structurally designed to serve clients more than workers."(P23)}
\end{quote}

Eleven participants also stressed that workers' agency is threatened by false or malicious reviews. On one hand, platforms do not provide legal protection against defamation. On the other hand, Bangladeshi labor laws do not extend to online labor. This leaves workers vulnerable. Even when a review clearly harms reputations, platforms often refuse to intervene. Hence, P27 opined, 

\begin{quote}
\textit{``Clients can lie or exaggerate in reviews. Online labor is not protected by defamation laws or labor rights, and without legal pressure, platforms default to neutrality and avoid taking sides. Only introducing platform labor defamation laws and reputation protection regulations can hold clients accountable for malicious reviews.", (P27)}
\end{quote}

\subsubsection{Ownership, Recognition, and Attribution Rights}
Seventeen participants felt that platform policies disregard their work, noting that a lack of required client acknowledgment often means their contributions go unrecognized and their portfolios suffer as a result. We noted 14 cases where clients purchased participants' works but imposed strict restrictions on referring to them in portfolios or conversations. Six participants were pressured to sign non-disclosure agreements (NDAs). Nine were offered extra money for remaining anonymous, while P4's and P1's pleas for acknowledgment were ignored. Eight participants faced threats of legal action for including previously completed tasks in their portfolios.

\subsection{Specific Contextual Challenges}
\subsubsection{Regarding Tasks with Questionable Moral and Ethics}
Often, the workers come across job postings with questionable morals and ethics; however, only workers are held accountable for taking and completing them. For instance, P7 accepted a job which later turned out to be a university assignment. Although he was aware that helping with assignments might raise ethical concerns, he assumed the listing met platform policies. He got his account banned upon finishing it,   

\begin{quote}
\textit{``I once accepted a task that turned out to be a university assignment. I completed it, but then my account got banned because their policy says you are not allowed to do academic work for students. I lost \$45 because of that. But if it was against the rules, then why did they allow that job to be posted on the platform at all? The fault was not mine. The platform should have filtered it.''} (P7)
\end{quote}

Similarly, P8 accepted a job involving a Python assignment. Although he was aware that helping with assignments might raise ethical concerns, he assumed the listing met platform policies. He got his account banned upon finishing it. Even after explaining, the platform did not reinstate it. The participant was frustrated that he was punished while the platform and job poster should've been held accountable.

\subsubsection{Geographic and Racial Hurdles}
Participants shared how platform laborers from certain countries in South Asia face more barriers, including account bans, limited onboarding, restricted services, and heavier scrutiny. For example, P13 described how \textbf{geographic relocation} triggered a false positive in the platform's monitoring system. After moving abroad, he created a new account since his original one had stopped functioning properly. However, the algorithm flagged this as an attempt to run multiple accounts and banned both profiles.

\begin{quote}
\textit{``I already had an account back in my home country. But after I moved to Japan for studies, my old account was not working. So I created a new account. Immediately, the new account was suspended. I think the algorithm assumed I was one person trying to operate multiple IDs, even though my old account was useless. In the end, both my accounts got suspended.", (P13)}
\end{quote}

P1 recalled a time when \textit{Upwork} had \textbf{access restrictions} and did not allow Bangladeshi users to even create accounts. Even after registration, Bangladeshi users often experience additional manual reviews to activate or enjoy their account functions because their location is Bangladesh. As he explained,

\begin{quote}
\textit{``There was a time when Upwork did not allow Bangladeshi users to sign up. Even when it did, there were often restrictions, and people had to go through manual reviews to make the account usable. During the time when opening an account was completely blocked, many people had to use VPNs to create their profiles.'' (P1)}
\end{quote}

Additionally, P3 highlighted that bans and restrictions seem to disproportionately affect users from South Asia. Based on his regular observations of online forums and discussions, most complaints about unexplained account deactivations or failures to recover access came from users in countries like Bangladesh, India, or Pakistan.

Strong \textbf{racial discrimination in wages and communications} was also discussed by 12 participants. Many clients are discriminatory against workers from the Global South, particularly South Asia, and offer unfairly lower payments and mistreat workers.  P4 shared his experience that clients, upon finding him to be from the ``Third World'',  decide that he deserves less compensation and wages than others. He told us about one of his bitter conversations,

\begin{quote}
\textit{``International clients commonly assume that in Third World countries, labor is valueless, and they decide to lower the payment too much. So, for example, they might get a \$50 job done here for just \$10. They would pay more to the workers from the other regions for the same task.'' (P4)}
\end{quote}

P12 described how this regional bias often goes beyond wage and becomes disrespectful. He recounted being treated rudely and harshly by clients who seemed to look down on South Asian workers. He believed clients' mistreatment persists because workers are not likely to be defended or protected by the platform.

\begin{quote}
\textit{``We are often looked down upon. Many times, clients behave rudely or speak harshly upon finding out I am South Asian. They treat us disrespectfully because they know that no one will stand up for us.'' (P12)}
\end{quote}

\textbf{Cross-border legal ambiguity} in platform work was another challenge the participants discussed. Global platform labor often operates in legal grey zones, where the boundaries of applicable labor laws and contractual expectations are unclear. Workers do not have formal employment agreements and rely solely on platform terms, which rarely clarify which country's labor standards apply. This leads to confusion and conflict when clients from one legal jurisdiction impose expectations that do not align with the worker's local context or legal protection. P11 shared an experience when a German client called him out for taking the day off during a religious festival. This upset both parties, as he described,

\begin{quote}
\textit{``(The German client) assumed I was legally required to follow their European work regulations. They expected me to respond within German working hours. I took a day off for Eid, and they got upset and accused me of violating `contract terms under EU rules.' But as a Bangladeshi, I am not bound by European labor law, and the platform never clarified whose legal standards apply.'' (P11)}
\end{quote}

\subsubsection{Abuse of the Absence of Fintech Platforms}
Nineteen participants pointed out that cross-country currency-related issues impact different hidden charges and platform fees, which reduce the actual earnings of the workers. These deductions often go unnoticed or unchallenged because they are built into the platform's non-negotiable terms of service. Fourteen participants explained that other costs, like unfavorable currency conversion rates, bank transfer fees, and digital taxes, also reduce workers' earnings. Absence of global payment systems such as \textit{PayPal} and \textit{Stripe} in Bangladesh made it more complicated for the participants. Hence, workers must rely on third-party intermediaries or workaround methods, which often introduce additional hidden costs and reduce transparency as the dollar's value fluctuates over time. Therefore, the challenges of ensuring fair wages in this space are bigger than expected, as P29 elaborated, 

\begin{quote}
\textit{``Platforms promise fair wages, but cannot deliver, as we do not even have renowned mediums like PayPal. Workers here are forced to go through agents or third parties who eat into the payments. This requires national FinTech reforms, global remittance policy negotiation, or platform partnerships with localized financial tools to solve these issues.", (P29)}
\end{quote}

Thus, ensuring fair wages is not just a platform concern, but there are other stakeholders in the space whose refusal to take accountability frustrated many participants.

\subsubsection{Conflicts with Algorithm}
Account getting erroneously suspended was one of the major frustrations for the participants. Twelve participants described being penalized or permanently banned by the platform's automated systems without sufficient explanation or human review. Handling such situations was particularly challenging, as their requests for clarity and appeal were addressed poorly by the platforms. For instance, P22 became a victim of \textbf{automated flagging} for having seemingly inappropriate words in submitted content. The project was about the developed chatbot being able to detect and take necessary actions on the use of inappropriate words, so the file included sensitive sample phrases used for testing the chatbot's responses. The platform's algorithm flagged it as inappropriate and penalized him without any contextual understanding or manual review.

\begin{quote}
\textit{``I uploaded the project files, some of which included sensitive phrases, to see if my tool can capture them. My account got flagged immediately. The platform algorithm picked up on the keywords and assumed I was doing something inappropriate. But testing with inappropriate words was the whole point of the testing. If a human had looked at the context, they would have easily understood it.", (P22)}
\end{quote}

There were even cases where uploading even ordinary, harmless files led to an unjustified account ban, as P7 shared. After completing a programming project, he tried to send the source code to the client. However, the platform's algorithm flagged with an unspecified violation, triggering a suspension. Despite repeated attempts to clarify the situation, his account was never restored.

Eight participants also highlighted a key challenge to algorithmic fairness. Platforms increasingly rely on \textbf{monetized visibility systems} where workers must pay to improve their ranking or exposure. This pay-to-play model distorts merit-based evaluation and disadvantages workers who cannot afford to promote themselves, regardless of their performance quality or reliability. Even workers with excellent track records are often overlooked in search results unless they invest in platform-sponsored boosts or ads. P27, among others, saw this as a threat to workers' equitable access to work and demanded better platform policies so that people seeking paid visibility would not sideline the actual skilled workers. 

\begin{quote}
\textit{``Workers with good performance are still invisible unless they spend money. This favors richer workers over poorer ones. I understand that paid exposure boosts platform revenue. However, policies must regulate algorithmic pay-to-play structures and revise fair visibility rules that cannot be bypassed by money.", (P27)}
\end{quote}

\subsection{Situated Solutions and Methods}
\subsubsection{Strategic Identity Framing to Bypass Regional Discrimination}
A form of situated knowledge that platform workers develop through community training and peer learning involves strategically framing their identity to gain entry into platforms. Major platforms like \textit{Fiverr} and \textit{Upwork} have strict policies that prohibit anyone under the age of 18 from creating an account. To address this, \textbf{borrowing ID} was a common practice in the community, particularly for young laborers who started early. The participants said they used some adult person's NIDs to seem eligible by age and to create accounts. P4 described he used his older brother's ID:

\begin{quote}
\textit{``When I first tried to open an account, I could not do it because I was under 18. Later, one of my trainers from the place where I was taking courses told me that if I could manage someone's NID who is over 18, I could easily open an account using that. My elder brother allowed me to use his NID, and I opened the account in his name and started working.'' (P4)}
\end{quote}

\textbf{\textit{Posing as a `Westerner'}} was another situated method they used to bypass regional discrimination. Clients tend to pay more and show greater respect when they believe a worker is based in the U.S. or Europe, even if the quality of work is identical. Over time, workers built informal knowledge networks around how to ``look Western" through naming strategies, fake location tags, VPNs, or remote desktops. By switching location tags or adjusting linguistic style, they discovered more opportunities and higher pay. Participant P31 described how he used a trusted friend's identity to create the impression of being U.S.-based:

\begin{quote}
\textit{``My friend living in the USA helped me open my account. First, VPN was troubling, so I used his remote desktop to make it seem like I was based there. I used his phone number for verification and provided his US bank details. Basically, he became the face of the profile, but I was doing all the work behind. That is how I managed to create an American profile. Huge difference: I started earning way more than what I could have earned with a Bangladeshi profile.'' (P31)}
\end{quote}

Another method we noted was \textbf{profile revision}. When too many platform laborers offer the same type of service, platforms become highly selective, rejecting new accounts that do not appear exceptionally strong or unique. As a workaround, workers chose less saturated skills during account setup, just to get the profile approved. Once inside the platform, they later revised their profile and switched to their actual area of expertise. P12 explained how he passed the gate-keeping:

\begin{quote}
\textit{``...(T)hey kept rejecting it again and again. Even though my profile was complete and everything looked fine, the platform just would not accept it, but they flagged it every time. Then my trainer told me that maybe the platform was oversaturated with logo designers. He advised me to list a different skill, like something more uncommon, and try again. I followed that advice, picked a less common skill, and then the account finally got approved.'' (P12)}
\end{quote}

This set of tactics reflects a deeply postcolonial reality where workers must erase or obscure their true identity and attributes to be treated fairly through algorithmic scrutiny. This curated and embodied resistance against the systematic process of denial to acknowledge the Global South laborers' existence reveals how situated knowledge emerges from persistent inequality and survival within digital infrastructures.


\subsubsection{Sneak in Contact Information to Bypass Platform Fees}
Most major platforms (\textit{Fiverr, Upwork}, etc) deduct a 20\% fee from every transaction, regardless of the worker's earnings. For many workers in the Global South, where payment rates are already far lower, this fee becomes a heavy burden. Over time, through trial and error, workers developed creative methods to shift clients off-platform, where payments could be made directly without deductions. A common tactic among platform workers was \textbf{contacting clients privately}. However, platforms use automated systems and content-detection bots to monitor chats and block any message that seems to include contact information such as phone numbers, emails, or links. In response, workers have learned to \textbf{encode their off-platform contact information} subtly by embedding it inside images, breaking up numbers, or using indirect language. This situated knowledge helps workers retain more of their already-meager income and resist the platform's structural dominance. Participant P28 described a coded messaging style he developed for this purpose,  

\begin{quote}
\textit{``I send a long message that hides my WhatsApp number inside the regular sentences. I can write something like: I am working on one idea that needs to be shaped. You remember the seven things we discussed last time, and I have five cool tips to solve those. You can see that there are one, seven, and five already planned out.' So it looks like a normal message, but it actually means 0175.'' (P28)}
\end{quote}

Alongside linguistic tricks, we also noted that others used \textbf{visual techniques} to escape detection. A different workaround involves embedding contact details within images rather than in the text box. Workers realized that the platform's bots are trained to scan written messages, but often cannot read text hidden within graphics. Using this loophole, some workers now share screenshots of their work with their contact information disguised in a corner or styled to look like part of the project itself.

\begin{quote}
\textit{``My trick is sending updates as screenshots. Like, if I am reporting the progress of a design or some code, I sometimes add my email or WhatsApp number in a small corner of the image with different symbols, maybe in the footer or next to a fake label, so the client notices it, but the platform bots can not catch it. It is risky, but if the client really wants to work directly, they usually respond.'' (P23)}
\end{quote}

These examples reflect a situation where workers from the Global South must conceal themselves and tactically bend platform rules to retain economic value from their own labor. This knowledge emerges from lived experience, community sharing, and a persistent need to survive within unequal platform labor infrastructures.

\subsubsection{Tactical Software Use and Automation Shortcuts}
Platform workers learned how to manipulate software tools in unconventional ways to reduce effort, avoid costs, or bypass restrictions. These tactics are commonly passed down through informal training centers, online groups, or peer-to-peer mentoring. One major area of adaptation was using pirated or cracked software. For example, licensed software like Adobe Photoshop is unaffordable to many Bangladeshi workers. In response, a widespread workaround has emerged: using older, more stable cracked versions that are known to work reliably without detection. Trainers often provide lists of ``safe'' versions that should not be updated, since newer releases introduce security patches that break the cracked copies. Over time, workers develop a shared understanding of which software setups are functional and which ones to avoid. P33 explained that he was trained to avoid updating his pirated copy of software to ensure uninterrupted work:

\begin{quote}
\textit{``...(W)hen we used those, we had to be careful not to update them. Newer versions usually have higher security features, and sometimes they cannot be cracked. They even gave us a list of which versions of which software are safe to use and which ones should not be updated. For example, we were told to always use Photoshop 2018, because its cracked version is easily available and stable.'' (P33)}
\end{quote}

Another area where situated knowledge thrives is in task automation. Workers often devise multi-tool workflows that reduce labor time while appearing as if the work was done manually. These workflows are circulated within peer communities. They combine public services, platform loopholes, and free online tools into streamlined pipelines. For example, transcription workers may rely on YouTube's automatic captioning as a hidden backend by treating YouTube not as a platform for content, but as an invisible transcription engine. P26 shared how she learned to use automated tools creatively to cut down on transcription time:

\begin{quote}
\textit{``...(S)o that we do not have to manually type everything out. Instead, we upload the audio or video privately on YouTube and let it generate auto-captions. Then we download those captions and use an online tool to fix grammar or make small edits. After that, we send the polished version to the client. Earlier, a task like this would take me two to three hours, but now I can do it in 20 to 30 minutes.'' (P26)}
\end{quote}

These examples reveal how software is not merely a tool for work but also a space of negotiation, curation, and workaround. Workers build intimate, sometimes subversive knowledge of how digital systems can be stretched, broken, or quietly repurposed to serve their needs. These practices reflect a postcolonial computing reality where access, affordability, and global inequality shape what tools people use and how they learn to survive through them.

\subsubsection{Boosting Ratings Through Volume of Small Tasks}
Another situated method involves accepting many small, low-paying tasks to boost ratings. On most platforms, visibility and credibility depend on average star ratings and the number of completed projects. Even a single low rating can harm future chances, so workers strategically complete quick jobs to accumulate consistent 5-star reviews. Over time, it becomes their routine. P19 recalled being trained to focus on small jobs specifically to build a strong profile:

\begin{quote}
\textit{``My trainer suggested taking as many tasks as possible, no matter how low the payment is or how short the task is. Because the more small jobs I complete, the more 5-star ratings I collect. My overall rating will keep increasing. Then later, I'll be more likely to get bigger and better-paying projects. My profile will also be more visible to clients.'' (P19)}
\end{quote}

Thus, the strategy of accepting small and low-paid tasks is treated as an investment for obtaining a stack of high-rated reviews. A high rating makes workers more likely to secure larger, better-paying projects later.

\section{Discussion}
This paper examined how Bangladeshi platform workers enter digital labor markets, cultivate expertise through situated practices, confront algorithmic inequities, and develop tactical repertoires for survival. Our study extends and advances the arguments of Gray and Suri's Ghost Work \cite{gray2019ghost}, which revealed how hidden human labor sustains digital platforms while remaining systematically obscured. While their work emphasized the erasure and invisibility of workers' contributions, our findings demonstrate that sustaining AI systems also depends on what we call ghostcrafting: the situated, improvisational practices through which workers materially keep both their livelihoods and the AI supply chain functioning. Bangladeshi workers did not simply disappear behind interfaces but actively crafted survival through repertoires such as posing as Westerners with VPNs and borrowed IDs, embedding WhatsApp numbers in images to bypass fees, relying on pirated “safe” software versions, or sustaining careers from rented cyber-café computers. These tactics show that ghost work is not only hidden but also crafted in fragile infrastructures shaped by Global South constraints such as fintech exclusions, linguistic marginalization, and geographic bans. By documenting learning ecologies, communal infrastructures, and tactical survival, we move beyond diagnosing invisibility to articulating ghostcrafting as a general condition of contemporary AI production, one that demands recognition, redistributive policies, and design interventions rather than symbolic acknowledgment alone. The lessons from this research offer several immediate and broader implications and open discussion in both the HCI-design and theory ends. 

\subsection{Implication to Design}
We open the discussion by focusing on the design implications that our study generates for HCI and HAI. We found that rating and review systems often become tools of client control, with a single malicious or arbitrary review reducing workers' visibility and long-term opportunities. Participants repeatedly described how one unfair review could destroy their profile, even when clients left positive written comments. This resonates with prior work on the absence of accountability in digital labor platforms and the disproportionate power of algorithmic metrics \cite{13-grohmann2021beyond, 120-grossman2018crowdsourcing, 121-zou2018proof}. Analogous systems in other domains already offer workable remedies. For example, Uber's delayed mutual rating system and Coursera's moderated peer review panels reduce retaliatory or biased feedback \cite{kulkarni2013peer, rosenblat2017discriminating}. Translating these insights into design, we can introduce rating dashboards where workers log evidence of completed tasks, blind mutual review submission, and independent arbitration boards that restore visibility or income when governance fails. In a different subsection, we reported how language barriers systematically exclude Global South workers by making English or European languages the default. Participants like P2 and P13 emphasized how weak English proficiency or the absence of translation support caused them to lose clients and miss contracts. This aligns with scholarship that highlights how marginalized workers are disadvantaged by one-size-fits-all platform designs \cite{hcomp-flores2020challenges, 41-abbas2022goal}. Other fields provide models of inclusive infrastructure: Microsoft Teams supports live captions and translations across dozens of languages, while telehealth systems integrate real-time interpreters to bridge communication gaps \cite{Microsoft_Teams_LiveEvent_Captions, lu2020navigating, sharma2023language}. Drawing from these precedents, gig platforms could embed translation APIs into chat, offer customizable glossaries for task-specific terms, and provide optional human-mediated translation services. These interventions would potentially expand access to cross-border contracts, reduce exclusions rooted in language hierarchies, and recognize linguistic equity as a central design obligation.  

\subsection{Broader Implications}
Our research also provides some broader implications for HCI and HAI theory and policies. Below we discuss them under four themes: ghosted contributions and recognition, algorithmic domination and precarity, situated learning as expertise, and tactical repertoires for sustainable futures.

\subsubsection{Ghosted Contributions, Data Labor, and Recognition}
We contribute to invisible digital labor and data appropriation scholarship that discusses wage suppression, recognition gaps, and biased task allocation \cite{10-irani2023algorithms, 109-haralabopoulos2019paid, 55-salminen2023fair, 117-zhang2017consensus, 119-eickhoff2018cognitive, 13-grohmann2021beyond, 116-ford2015crowdsourcing}. Echoing concerns about unconsented data labor and synthetic data practices \cite{faact-li2023dimensions, faact-whitney2024real}, we show how NDAs, portfolio bans, and threats of legal action render workers' authorship unclaimable when their outputs train, tune, or maintain AI systems. This is not only a matter of credit; it forecloses career mobility, suppresses bargaining power, and obscures the human provenance of ``automated'' capability. Positioning our contribution relative to Gray and Suri's \textit{Ghost Work}, we conceptualize \textit{ghostcrafting AI} to foreground that invisibility is coupled with creative, situated, and continual \textit{crafting} of AI by workers who must learn, adapt, and repair under constraint \cite{gray2019ghost}. The value that accrues to AI systems is inseparable from this ongoing human craft---yet current contractual and platform arrangements strategically erase it. Therefore, we argue that recognition must be operationalized as \textit{attributable authorship pathways} in platform and client contracts (e.g., portfolio-safe fragments, time-bounded NDA carve-outs, standardized contributorship badges), paired with provenance-aware AI documentation, so that ghost contributions are institutionalized, visible, and career-productive.

\subsubsection{Algorithmic Domination and the Politics of Precarity}
We also contribute to algorithmic management, power asymmetries, and the gap between promised flexibility and lived precarity discussions in gig economies scholarship \cite{gig-spreitzer2017alternative, gig-watson2021looking, 66-wu2024gig, gige-donovan2016does, 58-stewart2017regulating}. In line with work showing how platforms privilege efficiency over fairness \cite{lit4-kuhn2021human} and exert tight control through ratings and rules \cite{lit1-hickson2024freedom, lit3-gerber2021community}, we document overpriced certification schemes, biased visibility algorithms that penalize inactivity or minor rating changes, and review mechanisms that disproportionately shape reputations and future access to jobs. These mechanisms produce path-dependent careers, where a single downturn compounds into invisibility, and they make exit costly for workers who depend on platform income. Recent scholarship urges worker participation and feedback as potential correctives, while warning against extractive or tokenistic models \cite{eaamo-sloane2022participation, eaamo-birhane2022power, eaamo-corbett2023power, eaamo-russo2024bridging}. Our data suggest that participation without governance teeth will not counteract the structural leverage embedded in rating, ranking, and certification markets. Participation must be coupled with enforceable rules for auditability, contestability, and due process. Borrowing insights from proposed accountability infrastructures in machine learning pipeline \cite{eaamo-barker2023feedbacklogs}, we argue that algorithmic accountability in platform labor must move from disclosure-only to \textit{actionable governance}, which mandates appeal channels, independent audits, and non-retaliatory remedies that directly constrain rating, ranking, and credential markets that manufacture precarity. 

\subsubsection{Rethinking Expertise: Situated Learning Beyond Formal Infrastructures}
Our findings join the ongoing discourse on how crowd workers learn and adapt in resource-constrained settings, extending work that documents ethical blind spots in management and training \cite{136-durward2016there, 137-bhatti2020general} and how tools, interfaces, and communities shape learning opportunities \cite{138-drechsler2025systematic}. Whereas prior studies surface informal and peer-led learning among rural and marginalized workers \cite{hcomp-flores2020challenges, 41-abbas2022goal, hcomp-imteyaz2024human}, our ethnography details concrete, situated practices that make digital work possible in Bangladesh: renting cyber-café computers on shifts, copying high-performing gig descriptions, following tutorials in unfamiliar languages, and troubleshooting in WhatsApp groups. These practices constitute an improvised pedagogy that substitutes for absent institutional pathways and reframes ``upskilling'' as a continuous, collective craft under infrastructural scarcity. Building on reviews that call out the under-attention to workers' own perspectives and conditions \cite{136-durward2016there}, we show how workers actively assemble local infrastructures of learning to meet global quality expectations. This assembly work does more than keep individuals employable; it is foundational to the stability of platform markets and the AI pipelines that depend on them. Therefore, we argue that HCI and HAI must recognize and design for \textit{situated learning ecologies}, including peer networks, shared templates, and localized tutorials, as first-class infrastructure for platform labor rather than treating training as an individual responsibility or an afterthought of task design.

\subsubsection{Tactical Repertoires and Designing Sustainable Futures}
Our findings join the ongoing discourse on accountability gaps and the long-term sustainability of platform work \cite{120-grossman2018crowdsourcing, 121-zou2018proof, 16-mcinnis2016running, 13-grohmann2021beyond, 17-mcinnis2017crowdsourcing, 25-naude2022crowdsourcing}. Prior work shows repetitive short-term tasks, unstable income, and burnout, alongside weak oversight and rare implementation of proposed audits or feedback logs \cite{111-berg2015income, 55-salminen2023fair, eaamo-barker2023feedbacklogs, 46-de2022understanding}. We add a granular account of \textit{survival infrastructures}: identity masking to reach better-paying markets, fee-bypassing via embedded contact info, stabilizing careers with ``safe'' pirated software, and visibility boosting through many low-pay tasks. These tactics are not merely rule-bending; they are repair practices that keep workers, and by extension the AI supply chain, functioning amid institutional absence. As Widder notes, fairness, accountability, and sustainability are often treated separately, obscuring their interdependencies \cite{faact-widder2024epistemic}. Our analysis shows how unfair ratings/visibility regimes (fairness) endure without oversight (accountability) and push workers into high-cost tactical repertoires (sustainability), which, in turn, entrench informality and risk. Design responses limited to UX tweaks or voluntary ``best practices'' will not realign these dynamics. Therefore, we argue that sustainable futures require \textit{stacked interventions}, including binding accountability (independent audits and grievance resolution SLAs), redistributive fairness (floor rates with hidden labor and demotion caps), and capacity infrastructures (localized training funds and portfolio-safe attribution channels). These measures ensure that workers' tactical repertoires become optional complements rather than compulsory survival.

\subsubsection{Postcoloniality in AI Labor Market}
Our work joins the ongoing postcolonial influence in computing and AI discourse \cite{philip2012postcolonial, irani2010postcolonial}. We highlight how Bangladeshi platform workers confront structural asymmetries that are deeply postcolonial in character. Geographic restrictions, racialized wage hierarchies, and linguistic exclusions reproduce colonial-era logics of devaluing labor from the Global South while centering Euro-American norms as defaults. Workers’ repertoires, such as posing as a Western person with VPNs or encoding WhatsApp numbers to avoid fees, are not simply technical tricks but forms of survival within infrastructures designed against them. Prior scholarship shows the mismatch of AI labor exploitation in the Global South \cite{regilme2024artificial, 13-grohmann2021beyond, 22-png2022tensions, 23-okolo2023addressing}, and our study extends this by showing how workers must tactically remake identities and infrastructures just to remain visible and employable. For HCI and HAI, this underscores that platform design cannot be understood apart from histories of coloniality: treating Bangladeshi workers merely as peripheral data laborers obscures how their craft sustains AI systems while reproducing inequities. Recognizing postcolonial dynamics requires moving beyond inclusion rhetoric toward redistributive design and governance that confronts these global asymmetries directly.

\subsection{Limitations and Future Work}
Our study focuses on Bangladeshi platform workers and, while offering rich ethnographic insight, may not generalize to all Global South contexts or to other sectors of gig work. The participant pool captures diverse entry points and experiences but remains limited in scale relative to the breadth of the industry. While biography making and interviews allowed us to surface lived experiences, our analysis is constrained by participants’ willingness to share and by our positionality as researchers, which may have shaped interpretation. Despite these limitations, our work is important because it foregrounds the lived experiences of Bangladeshi platform workers, offering grounded insights into how the people of global AI systems are materially sustained at the margins. Our future work would expand to comparative studies across multiple countries to better understand how structural gaps vary regionally. We also see opportunities for longitudinal research that tracks how workers’ survival repertoires evolve as AI platforms and policies shift.

\section{Conclusion}
This paper examined Bangladeshi platform labor through biographies, challenges, and situated survival practices, conceptualizing \textit{ghostcrafting AI} to describe how invisible and precarious human craft sustains global AI systems. We showed how workers enter through disrupted educational and economic pathways, build expertise in patchwork infrastructures, and navigate algorithmic precarity with tactical repertoires. Our findings highlight overlooked forms of identity struggles, communal learning, and survival mechanisms. We call on HCI and HAI to design infrastructures that recognize, redistribute, and sustain labor at the margins.


\bibliographystyle{ACM-Reference-Format}
\bibliography{sample-base}

\end{document}